# Interfacial spin-orbit-coupling-induced strong spin-to-charge conversion at an all-oxide ferromagnetic /quasi-two-dimensional electron gas interface

Mi-Jin Jin[1,2,*], Guang Yang[2,3], Doo-Seung Um[4], Jacob Linder[5], and Jason W.A. Robinson[2,*]

[1] Center for Multidimensional Carbon Materials (CMCM), Institute for Basic Science (IBS), Ulsan 44919, Republic of Korea
[2] Department of Materials Science & Metallurgy, University of Cambridge, 27 Charles Babbage Road, Cambridge CB3 0FS, United Kingdom
[3] School of Integrated Circuit Science and Engineering, Beihang University, Beijing 100191, China
[4] Department of Electronic Engineering, Jeju National University (JNU), Jeju-do 63243, Korea
[5] Center for Quantum Spintronics, Department of Physics, Norwegian University of Science and Technology, NO-7491 Trondheim, Norway

*Corresponding Author :
Mi-Jin Jin : mijin8276@gmail.com, jinmijin@ibs.re.kr
Jason Robinson : jjr33@cam.ac.uk

**Keywords**
Spin-Orbit Coupling, Spin Hall effect, Spin-Charge conversion, Oxide Interface, ferromagnetic resonance

**Abstract**

Functional oxides and hybrid structures with interfacial spin orbit coupling and the Rashba-Edelsterin effect (REE) are promising materials systems for thermal tolerance spintronic device applications. Here, we demonstrate efficient spin-to-charge conversion through enhanced interfacial spin orbit coupling at the all-oxide interface of $La_{1-x}Ca_xMnO_3$ with quasi-two-dimensional (quasi-2D) $SrTiO_3$ (LCMO/STO). The quasi-2D interface is generated via oxygen vacancies at the STO surface. We obtain a spin-to-charge conversion efficiency of $\theta_\parallel \approx 2.32 \pm 1.3$ nm, most likely originating from the inverse REE, which is relatively large versus all-metallic spin-to-charge conversion materials systems. The results highlight that the LCMO/STO 2D electron gas is a potential platform for spin-based memory and transistor applications.

**Introduction**

In spintronics, the spin Hall effect (SHE) originates from spin orbit coupling (SOC) plays an important role in spin-to-charge interconversion, magnetization switching, and spin current manipulation [1-20]. Nevertheless, the SHE and inverse SHE are a bulk phenomena with 3-dimensional propagation of spin. Recent reports have shown that the interfacial SOC effect from Rashba-Edelstein effect (REE), inverse Rashba-Edelstein effect (IREE) or the spin Galvanic effect (SGE) are key for spin and/or charge generation and detection in low dimensional systems [1, 21-28]. Rashba SOC arises due to spatial broken symmetry at surfaces and interfaces, lifting spin degeneracy and causing locking between momentum and spin degrees of freedom. At symmetry broken interfaces, charge flow creates a non-zero spin accumulation (i.e., the REE).

$SrTiO_3$ (STO) has a highly tunable quasi-2D conductivity [29], intrinsic/extrinsic SOC [30], structural symmetry broken intrinsic Rashba interface [31], and ferroelectricity [22]. STO is, therefore, promising for artificially manipulating interface properties with oxide ferromagnetic layers – see, e.g., studies on oxide magnetic thin films including $La_xCa_{1-x}MO_3$ (LCMO), $La_xSr_{1-x}MnO_3$ (LSMO), and Yttrium iron garnet (YIG) [32-42]. However, studies on the interaction between such a magnetic oxide layer and conductive oxide surface layer remain challenging and topical since they are important for the development of spin-orbitronic and spin-interface electronics applications.

Ferromagnetic resonance (FMR) spin pumping is an established technique for probing magnetic dynamic properties of ferromagnetic materials including spin-to-charge conversion (or vice versa). Recently, spin currents generated by a REE-driven spin accumulation (which is IREE) has been reported using FMR with spin pumping in $CoFeB/LaAlO_3/SrTiO_3$ structures[43], and spin generated by charge on interfacial conducting layers in $NiFe/Al/SrTiO_3$ [44]. Charge generated by spin on ferromagnetic layers via FMR with spin pumping have also been reported in $Py/LaAlO_3/SrTiO_3$ [45], $NiFe/Al/SrTiO_3$ [21, 22], $La_{0.67}Sr_{0.33}MnO_3/LaAlO_3/SrTiO_3$ [31] heterostructures, and at $Bi_2Se_3$ [25], $Pt/NiFe$[2], and Graphene/yttrium iron garnet (YIG) [46] interfaces.

Here, we report efficient conversion of spin currents into charge currents via FMR of LCMO on STO and simultaneous spin pumping at the quasi-2D STO interface. The spin-to-charge conversion efficiency caused by IREE is $\theta_\parallel \approx 2.32 \pm 1.3$ nm at 5 K, which is relatively large versus all-metal-based materials systems. Our results can lead to potential applications in oxide-based systems for low-power detection and generation of spin in non-magnetic systems.

## Results and discussion

**Basic properties of LCMO/quasi-2D STO.** A conductive quasi-2D interface is artificially created between (001)-oriented STO and LCMO using Ar plasma treatment. As shown in Figure 1a, the sample was fabricated with a 6-armed Hall bar device (500 μm channel width, 1500 μm channel length) using optical lithography, reactive ion etching and thermal annealing [29]. The LCMO has a thickness of 15 nm, determined by X-ray reflectivity, (XRR) (Figure 1b) using Gen X fitting software. We used $La_{0.7}Ca_{0.3}MnO_3$ source material for a Pulsed Laser Deposition (PLD). The X-ray diffraction (XRD) patterns (supporting information Figure S1) support the LCMO main growth direction mainly follows the STO (001) orientation. Atomic Force Microscopy (AFM) on the quasi- 2D surface (Figure 1b inset) shows that the LCMO surface is smooth with vertical roughness of a root-mean-square less than 1-nm over a 5 $\times$ 5 μm$^{-2}$ area. For more information, we added two different sample surfaces as a comparison depends on STO surface condition (see Figure S2 on supporting information). Volumetric magnetic properties of the LCMO are investigated using a vibrating sample magnetometer (VSM) at different temperatures and magnetic fields. From the VSM measurements, we can estimate that a dead layer of the LCMO shows less than 5 nm (Figure S4). In Figure 1c we have plotted the temperature-dependence of the magnetic moment (field cooling, FC) with an external magnetic field of 100 mT after Hall bar geometry device fabrication. The LCMO shows a rise in magnetic moment with decreasing temperature, reaching a maximum moment of 10 μemu at 2 K, from which we estimate a Curie temperature ($T_c$) of 175 K. Note that the LCMO magnetic properties and curie temperature are thickness-dependent [47]. The inset in Figure 1c shows magnetic moment vs in-plane magnetic field hysteresis loops for the LCMO Hall bar at 4 different temperatures. At 2K, the magnetic coercivity of the LCMO is about 100 mT.

Next, we investigate the quasi-2D interface of STO (after post deposition of LCMO and Hall bar pattern). The temperature dependence of the sheet resistance is shown in Figure 2, confirming that the quasi-2D interface of STO is stable with metallic behavior and a sheet resistance of ~ 50 Ω/□ at 2 K. Note that the LCMO layer shows semiconductor-like behavior and a different resistance range. Supporting Information Figure S3 shows the temperature-dependent sheet resistance of the LCMO. Comparing with Figure 2, the behavior in Figure S3 demonstrates that the quasi-2D conducting interface is stable with metallic behavior through the temperature range. From the Hall effect and sheet resistance, we estimated a carrier concentration of the quasi-2D interface of $n_s = BI/(q|V_H|) = 1/(q\mu_s R_s) \approx 1.63 \times 10^{13}$ cm$^{-2}$, and mobility $\mu_s = |V_H|/(BIR_s) = 1/(qn_s R_s) \approx 11715$ cm$^2$/V.s at 5 K. Here, $q$ is the electron charge ($1.602 \times 10^{-19}$ C).

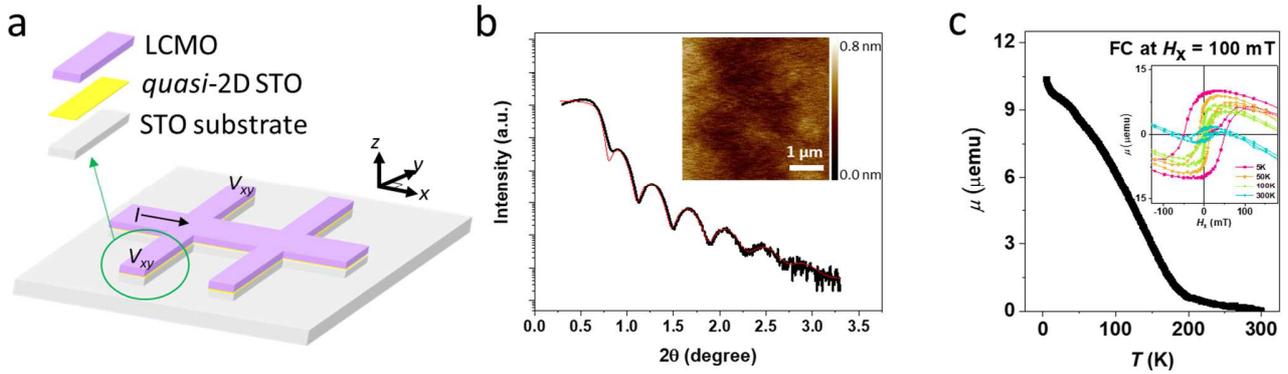

Figure 1. a. A schematic diagram of the (001)-oriented STO LCMO/STO Hall bar. b. X-Ray Reflectivity (XRR) of the LCMO/STO stack (red line is the fitted data). 2θ is angle between the transmitted X-ray beam and the reflected X-ray beam. Inset image shows the surface morphology of LCMO measured from an Atomic Force Microscopy (AFM). c. Magnetic properties measured using a vibrating sample magnetometer. Magnetic moment ($\mu$) vs temperature ($T$) at 100 mT (field cooling, FC) and magnetic moment ($\mu$) vs fieldsweep ($H_z$) at different temperatures [5 K, 50 K, 100K, and 300K] (inset).

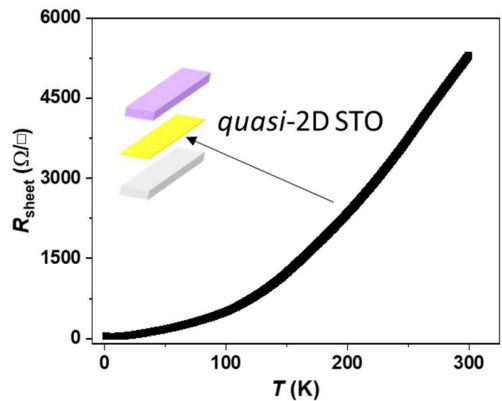

Figure 2. Sheet resistance ($R_S$) vs temperature ($T$) of the quasi-2D STO interface. $I_{source}$ = 1 µA. The STO conducting interface shows metallic behavior over the temperature range from 2 K – 300K.

**FMR with spin pumping_spin-to-charge conversion.** FMR spin pumping measurements are used for evaluating interfacial SOC [21, 43, 45] at a ferromagnetic / conducting layer interface. Figure 3c shows a schematic diagram of the FMR setup with spin pumping on the LCMO / quasi-2D STO / STO structure. Figure 3a shows a representative FMR signal from LCMO/STO at a microwave frequency of 8 GHz and power of 25 dBm. The LCMO shows a typical FMR response (Lamor precession) as the microwave voltage passes along the LCMO surface (Figure 3a). Simultaneously, a spin current from LCMO is injected into the interface, generating a nonequilibrium spin distribution. The resulting electron spin distribution leads to a voltage difference $V_{dc}$ across the direction parallel to the interface. (Figure 3b). The generated voltage by spin pumping likely arises due to the strong SOC at the LCMO/STO interface. Therefore, the spin pumping voltage (i.e., $V_{dc}$) is evidence of strong SOC at the quasi-2D interface. The FMR derivative can be fitted using $V_{\text{FMR derivative}} = V_{\text{symmetry}}F_{\text{symmetry}} + V_{\text{Asymmetry}}F_{\text{Asymmetry}}$, where $F_s$ and $F_A$ are the symmetric and antisymmetric Lorentzian functions, respectively. As shown in the inset of Figure 3a, the amplitude of the symmetric voltage ($V_{\text{symmetry}}$) and antisymmetric voltage ($V_{\text{Asymmetry}}$) in the inset of Figure 3a (red and blue fitted line) are correlated and are related to damping-like- and field-like-torque terms, respectively [43, 48]. The relatively large symmetric component of $V_sF_s$ suggests an in-plane spin polarization but not exact, damping-like torque is dominant. Also, the spin pumping signal ($V_{dc}$) can be analyzed for the symmetric and antisymmetric components, correlating SOC strength to spin pumping and anisotropic magnetoresistance, and a planar Hall effect [23, 49-51], respectively. A large symmetric component of $V_{s,dc}F_{s,dc}$ as shown in Figure 3b inset (red fitted line) suggests a large SOC-related spin pumping is dominant, even though an anti-symmetric component still exist [Figure 3b inset (blue fitted line)][2]. We separately plot the spin to charge conversion voltage $V_{s,dc}$ voltage as a function of temperatures in Figure 3d. The spin to charge conversion voltage is diminishes with increasing temperature. The original plots of LCMO FMR with spin pumping signal at 2 K (Figure S5a and b) and at 10 K (Figure S5c and d) are shown in the Supporting Information, part 4.

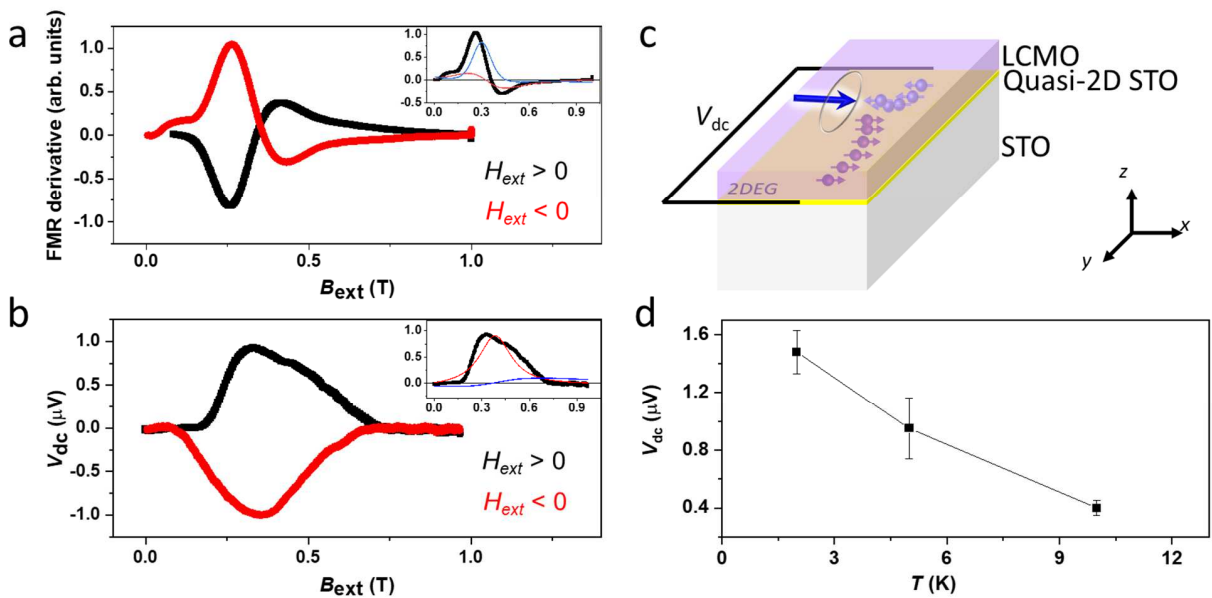

Figure 3. a. FMR signal from LCMO on (001)-oriented STO. (inset) FMR signal fitted with a symmetry component (blue) and an asymmetry component (red). b. The spin pumping signal detected via FMR. Spin-based pumping voltage signal detected from the conducting quasi-2D STO layer at positive and negative

external field. (inset) Spin-based pumping signal fitted with a symmetry component (blue) and an asymmetry component (red). The FMR peak and spin pumping measurement were taken at 5 K. c. A schematic illustration of FMR with spin pumping measurement at a LCMO / quasi-2D STO / STO. d. Temperature vs spin pumping voltage $V_{s,dc}$. The spin-to-charge conversion voltage gradually decreasing by increasing temperature.

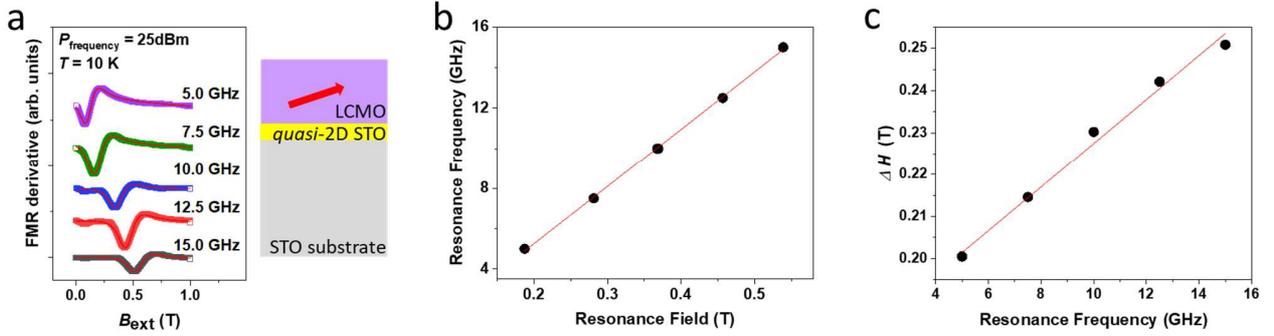

Figure 4. a. FMR signal from LCMO vs DC magnetic field with different applied microwave frequencies (labelled). Right illustration shows the cross-sectional structure of the device. b. Resonance frequency (GHz) vs resonance field, fitted using the Kittel's relation (red line) c. peak-to-peak linewidth $\Delta H$ depends on resonance frequency (GHz). From fitting of b and c plots we determine the characteristic values of the effective magnetization and Gilbert damping constant of the LCMO layer.

To investigate further the magnetic properties of the LCMO layer, we have measured the FMR signal versus FMR frequency. Figure 4a shows the dynamic magnetic resonance at several different frequencies $f$ of the microwave excitation, and versus external magnetic field $H_{ext}$. The corresponding frequency-dependence of the resonance field and the 1st derivative of the resonance peak-to-peak line width $\Delta H$ are reported in Figures 4b and 4c, respectively. The dispersion relation of $H_{res}$ with $f$ follows the Kittel relation $f = \frac{\gamma}{2\pi}[H_{res}(H_{res} + \mu_0 M_{eff})]^{1/2}$ [31, 52, 53] with $\gamma$ the gyromagnetic ratio, $H_{res}$ is the resonance magnetic field, and $M_{eff}$ is the effective magnetization (0.02 ± 0.01) kA/m. The effective Gilbert damping constant $\alpha_{LCMO/quasi-2D\ STO}$ is determined by fitting the frequency dependence of the peak-to-peak line width $\Delta H$ (Figure 4c) using $\Delta H = \Delta H_0 + \frac{4\pi \alpha_{LCMO/quasi-2D\ STO}}{\sqrt{3}\gamma}f$ where $\Delta H_0$ is the frequency independent contribution from magnetic inhomogeneous in LCMO. The Gilbert damping $\alpha_{LCMO/quasi-2D\ STO}$ is estimated to be about 0.19 ± 0.1 at 10 K.

The spin-to-charge conversion efficiency is related to SOC via the relation $\theta_{\parallel} = j_c^{2D}/j_s$. A quantitative analysis of spin-to-charge conversion requires the pumped spin current density to be determined using the relationship as follows: [21, 31, 52, 54]

$$j_s = \frac{g_{eff}^{\uparrow\downarrow}\gamma^2(\mu_0 h_{rf})^2 \hbar \left(\mu_0 M_{eff}\gamma + \sqrt{(\mu_0 M_{eff}\gamma)^2 + 4\omega^2}\right)}{8\pi\alpha_{LCMO/quasi-2D\ STO}^2\left[(\mu_0 M_{eff}\gamma)^2 + 4\omega^2\right]}\left(\frac{2e}{\hbar}\right), \qquad (1)$$

where $g_{eff}^{\uparrow\downarrow}$ is the effective spin mixing conductance $g_{eff}^{\uparrow\downarrow} = \frac{4\pi M_{eff} t_{FM}(\alpha_{LCMO/quasi-2D\ STO} - \alpha_{LCMO})}{g^*\mu_B}$ [21, 31], $\gamma$ is the gyromagnetic ratio, $h_{rf}$ the microwave field amplitude, $\omega = 2\pi f$, and $t_{FM}$ = 15 nm is the thickness of the LCMO film. From the above, we obtain average $g_{eff}^{\uparrow\downarrow} \approx 7.03 \times 10^{16}$ m$^{-2}$, which is reasonable versus values reported elsewhere at interfacial systems [55]. $j_s \sim 2.5 \times 10^4$ Am$^{-2}$. Which, in conjunction with the measured 2D charge current, results in a spin-to-charge current conversion efficiency of $\theta_\parallel = j_c^{2D}/j_s \approx$ 2.32 ± 1.3 nm. This value in our LCMO/quasi-2D STO/STO structure is relatively large versus metal based interfaces [53, 55-57], but also consistent with other oxide systems [21-23, 31].

**Discussion.** The spin-to-charge conversion at the quasi-2D interface most likely originates from IREE . An injected spin accumulation will generate a charge current flowing parallel to the interface, causing a measurable voltage across the ends of the sample. For the IREE case, the band structure is split due to the Rashba effect at the oxide interface. The spin current from the LCMO is injected into the LCMO / quasi - 2D STO interface at FMR, generating a charge current in the lateral (*y*) direction (see Figure 3c). This effect induces an electromotive force between the electrodes at the edges of the sample along long *y*. The electromotive force includes a signal from the LCMO that is induced by the microwave electric field through the galvanomagnetic effects (ex. Anomalous Hall effect and planar Hall effect) which should be separated from the IREE signal as shown in Figure 3. The possible origin of this IREE could relate to interfacial stemming from the broken inversion symmetry at the interface, or impurity-(such as oxygen vacancy) induced extrinsic interfacial spin orbit interaction [58-60]. Note that the sign of the spin pumping can sometimes be same value for opposite external magnetic fields. This behaviour may have two possible origins: one, if the ferroelectric polarity of STO is unstable [22]; two, a competition between spin and orbital Edelstein effects [61].

**Conclusion**

In this study, we have experimentally demonstrated a strong spin current to charge current conversion at an all-oxide interface between LCMO and STO using FMR spin pumping. In particular, we show strong spin-to-charge conversion via SOC at the quasi-2D LCMO/STO interface. From these results, we obtain a spin-to-charge conversion efficiency of $\theta_\parallel \approx 2.32 \pm 1.3$ nm which originates from interfacial SOC through the spin diffusion length. The value range is relatively large versus metallic based systems. Our results highlight that two-dimensional electron gases with functional oxide layers are strong candidates for spintronics and spin-orbitronic low power devices.

**Methods**

**Sample preparation**. 5 mm by 5 mm single crystal (100)-oriented (with < 0.1° tolerance from CrysTec GmbH) STO substrates were used. Substrates are cleaned by acetone, ethanol, and deionized water with sonication. Ar plasma treatment was followed to creating oxygen vacancies on the STO surface, which leads to conductive surface [29]. In-situ Ar plasma (50 W power) treatment is applied to the STO surface for 30 minutes at a pressure of 30 mTorr (Base pressure up to $10^{-9}$ torr). After then, the sample directly transferred to Pulsed Laser Deposition (PLD) system chamber and wait until base pressure of $10^{-6}$ torr range. Finally, magnetic oxide layer of $La_{1-x}Ca_xMnO_3$ (LCMO) is deposited onto STO. We used $La_{0.7}Ca_{0.3}MnO_3$ as the target material. For the LCMO deposition, laser pulse of 3 Hz with an energy density $\approx 2$ J cm$^{-2}$ were applied at 700 °C with oxygen pressure $1 \times 10^{-1}$ torr. This condition gives an approximate growth rate of 1 nm per minute.

**Device fabrications**. The simple Hall bar shaped geometry is chosen for not only applying longitudinal charge current but also generate a spin Hall effect induced transverse voltage detection. For the patterning, simply patterned shadow mask (channel width 500 μm, and channel length 1500 μm) is used for the buffer layer deposition. A 100-nm-thick Al layer is then deposited as a buffer layer. After then, controlled (working pressure 20 mTorr) oxygen plasma treatment (Adixen, AMS 100) followed by oxygen annealing (3 hours, 300 °C) is used to form the Hall bar patterned conductive STO [29, 62]. Finally, the Al buffer layer is removed by using base solution. Detailed fabrication processes are described in the previous work [29]. After patterning, Au (30 nm) / Ti (10 nm) layers were deposited (Thermal and e-beam evaporation) for contact electrodes and direct conducting STO interface contact.

**Low temperature Analysis Measurements** Siver paste or indium paste are used to connect between sample electrodes and the measurement equipment puck through copper wires. Temperature-dependent sheet resistance, and low temperature magnetic field sweep dependent signal detection related with magnetization switching were studied in liquid helium dewar system with a sourcemeter (Keithley 6221), (Keithley 2636) and nanovoltmeter (Keithley 2182). Spin pumping followed by Ferromagnetic resonance (FMR) measurement were studied using Physical Property Measurement System (Quantum Design) with a lock-in amplifier (SR830 ), AC sourcemeter (Keithley 6221) and nanovoltmeter (Keithley 2182).

**Supporting information**

X-ray diffraction analysis; AFM image of Sample surface; LCMO thin film's temperature dependent sheet resistance; Dead layer estimation; FMR and spin pumping data at 2 K, 10 K

**Author information**

**Corresponding Author**

*E-mail: Jason W.A. Robinson (jjr33@cam.ac.uk) and Mi-Jin Jin (mijin8276@gmail.com)

**Acknowledgements**

This research was supported by the Institute for Basic Science (IBS-R019-Y1) and the Basic Science Research Program through the National Research Foundation of Korea (NRF) funded by the Ministry of Education (2019R1I1A1A01063889). J.W.A.R. acknowledge funding from the EPSRC through the EPSRC-JSPS Core-to-Core Grant "Oxide Superspin" (EP/ P026311/1). J.L. was supported by the Research Council of Norway through Grant No. 323766 and its Centres of Excellence funding scheme Grant No. 262633 "QuSpin."

**Author contributions**

M.-J.Jin devised the project. M.-J.Jin fabricated the devices and performed the experiments and analysis. Guang Yang, Doo-seung Um assisted device fabrication and characterization. Jacob Linder provided theory support. M.-J.Jin and J.W.A. Robinson wrote and revised the manuscript with input from all authors.

**Competing interests**

The authors declare no competing interests

**For Table of Contents only**

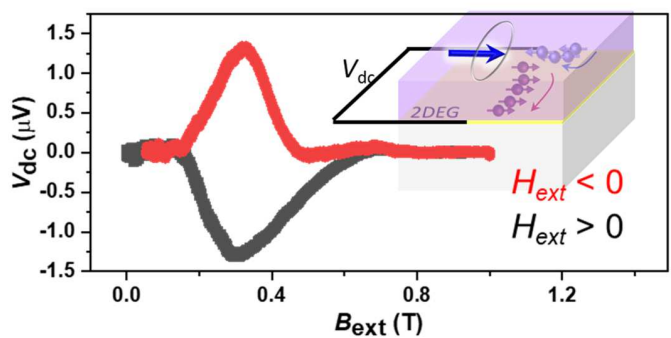